\documentclass[twocolumn,showpacs,preprintnumbers,superscriptaddress,amsmath]{revtex4}
\usepackage{graphicx}% Include figure files
\usepackage{dcolumn}% Align table columns on decimal \varepsilonpoint
\usepackage{bm}% bold math
\usepackage{amssymb}

\begin{document}
\title{Genetically engineered cardiac pacemaker: stem cells transfected with HCN2 gene and myocytes - a model}.

\author
{Sandra Kanani, Alain Pumir, Valentine Krinsky \\
{\small $^1$ Institut Non Lin\'eaire de Nice}, 
{\small 1361 route des Lucioles, 06560, Valbonne, France} 
}

\date{\today}

\begin{abstract}
 Artificial biological pacemakers were developed and tested in canine ventricles. Next steps will require obtaining oscillations sensitive to external regulations, and robust with respect to long term drifts of expression levels of pacemaker currents and gap junctions. 
We introduce mathematical models intended to be used in parallel with the experiments.

 The models describe human mesenchymal stem cells ({\it hMSC} ) transfected with HCN2 genes and connected to myocytes. They are intended to mimic experiments with oscillation induction in a cell pair, in cell culture and in the cardiac tissue. We give examples of oscillations in a cell pair, in a one dimensional cell culture and   dependence of oscillation on number of pacemaker channels per cell and number of gap junctions. The models permit to mimic experiments with levels of gene expressions that are not achieved yet and to predict if the work to achieve this levels will significantly increase the quality of oscillations. This  give arguments for selecting the directions of the experimental work.
\end{abstract}
 
\maketitle

{\bf Introduction} 

Genetically engineered pacemakers could be developed as a possible alternative to implantable electronic devices. 
Several approaches are developing: heterologous h2-adrenergic receptor expression in pig atria \cite{porcine2001}, over-expression of $I_f$ \cite{Ira2001} and dominant negative suppression of $I_{K1}$ in guinea pig ventricle \cite{Marban2002Nature}. 

Impressive results were obtained with stem cells to create an artificial biological pacemaker \cite{Rosen1}:
human mesenchymal stem cells ({\it hMSC}) transfected with HCN2 genes 
injected in canine left bundle branch provide ventricular escape rhythms that were mapped to the site of the injection. This experimental success shows great promises for future therapy.

In a recent review article \cite{Rosen}, the question : "What characteristics should be embodied in a biological pacemaker?" was put forward. The creation of a stable physiological rhythm was listed as one of the most important properties for future biopacemakers. The present article describes the mathematical models intended to be runned in parallel with the experiments to achieve this goal. 

The oscillations should be reasonably stable with respect to inevitable variations of the parameters affecting the biological pacemaker. Indeed, many factors can affect the period of the pacemaker namely: decreasing the expression levels of pacemaker channels or gap junctions, modification of the peptides regulating gene expressions, 
loss of genes, migration of small percentage of stem cells to other sites or differentiation into other cell types; loss of pacemaker function by some cells.
 
In order to obtain a stable pacemaker, many time consuming experiments should be performed. In exploring such systems, we believe that an adequate combination of experiments with theoretical predictions can lead to a much faster results, by reaching a better quantitative understanding and in so doing, in reducing
the number of experiments. To this end, one needs precise and adaptable mathematical models based on well-established concepts and results \cite{IraC1}-\cite{Moore}.

 The models presented here describe human mesenchymal stem cells 
({\it hMSC} ) transfected with HCN2 genes and connected to myocytes. 
They are supposed to be runned on a computer mimicking the planned experiments. A part of these computer simulations can be used instead of the planned experiments while experiments performed in parallel with the computer simulations can be used to update the models. 

 In the first version of the models, all the complex processes mentioned above are condensed into three control parameters: the number of pacemaker channels 
per stem cell, $N$; the number of gap junctions per cell, $n$ and the
concentration of the stem cells in the pacemaker.  
 
We demonstrate how they work by mimicking experiments with oscillations induction in a cell pair, in cell culture and in the cardiac tissue intended to study the robustness of oscillations and stability of the oscillation period.  
 
 The models permit to easily mimic experiments with arbitrary levels of gene expressions (including levels not yet achieved in  experiments) and to assess whether achieving such levels of expression will increase the quality of oscillations.
 
\vspace{.4cm}

{\bf  {\Large Models}}

%see also dir 2TextFor
\vspace{.4cm}

{\bf Stem cell models}

Models for a stem cell {\it hMSC} transfected with HCN2 gene were constructed from experimental data \cite{IraC1,IraPrivate}. 
We focus here on two models: 
a model, including only inactivation similar to models of HCN channels by \cite{DiFrancesco}, \cite{ZhangBoyett, modelKurata}, \cite{Marban05}  
and a model with two gating variables, activation and inactivation of the pacemaker current, similar to the Hodgkin-Huxley (HH) model \cite{HH}.

\begin{figure}[h]
\includegraphics[width=3.2in]{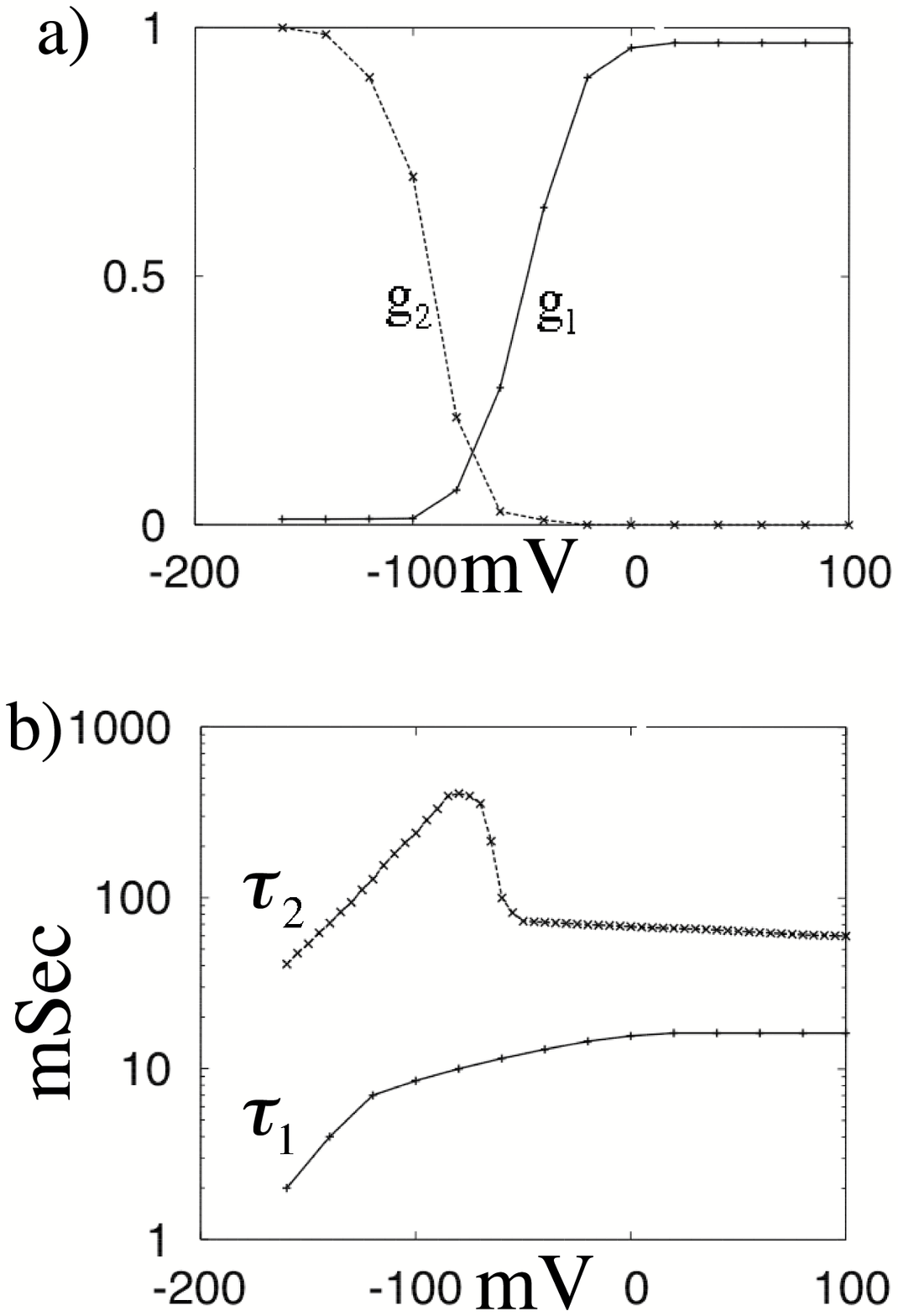}
\caption[] { {\bf  Stem cell model.}
a) steady values of activation $g_1$ and inactivation $g_2$. b) time constants.}
\label{g &tau_mh}
\end{figure}

%model

{\it  Stem cell model with inactivation only (Stem-h)} is:
\begin{subequations}
\label{Stem-h}
\begin{eqnarray}
d E_s /dt = -N  \sigma_f  \, \, g_2^s  \, \, [E_s - E_r] /C_s
\\
d g_2^s /dt=[{ \bar{g}_2^s(E_s)} - g_2^s] /\tau_2^s(E_s)
\end{eqnarray}
\end{subequations}

Here, $E_s$ is the stem cell membrane potential, mV.
$N$ is the number of pacemaker channels per cell, 
$\sigma_f~\sim 1 pS$ is the conductivity of a single pacemaker channel, C$\sim$95 pF is the stem cell capacitance, 
$E_r \sim$ - 40 mV is the reversal potential of the HCN2 pacemaker current. 

$g_2^s$ is a gating variable describing inactivation kinetics   
(notation $h$ is used in Hodgkin-Huxley eqns\cite{HH}). 
${ \bar{g}_2^s(E_s)}$ is its steady value that depends on $E_s$, 
$\tau_2^s(E_s)$ is the inactivation time constant that depends on $E_s$ as well. 

The model (Stem-h) is totally defined by two functions only,  
 ${ \bar{g}_2^s(E_s)}$, $\tau_2^s(E_s)$; their graphs are shown in   Fig.\ref{g &tau_mh}. 

%1
{\it Comment 1.} It is possible to use more detailed notations keeping track that the pacemaking current $I_f$ consists of two components, $I_{Na}$ and $I_K$ (e.g., as in \cite{ZhangBoyett,modelKurata}). 

\begin{subequations}
\label{Holden}
\begin{eqnarray}
d E_s /dt = I_{f,Na}Ê+ÊI_{f,K}Ê
\\
I_{f,Na}ÊÊ=Êg_{f,Na}   \, \, Y  \, \, [E_s-ÊÊE_{Na}]
\\
I_{f,K}ÊÊ=Êg_{f,K}   \, \,  Y   \, \, [E_s-ÊÊE_K]
\\
d Y /dt=({ \bar{Y}^s(E_s)} - Y)/\tau_2^s(E_s)
\end{eqnarray}
\end{subequations}

But since both components of the current pass through the same ionic channel and thus are gated by the same gating variable Y, this description collapses back to Eq.\ref{Stem-h}, with 

\begin{equation}
\label{ E_r}
E_r= [E_{Na} \, \,  g_{f,Na} + E_K  \, \,  g_{f,K} ]/[g_{f,K} +g_{f,Na}]     
\end{equation}

{\it Comment 2.} A model with one gating variable only (slow inactivation) is used in all descriptions of the HCN currents that we know. The justification of the above statement is that the pacemaker current induces oscillations with period $\sim$ 1 sec and thus there is no need for description of processes on a faster time scale.  

But this is not a reliable argument. The front of an AP induced by HCN develops on a much shorter time scale. To accurately describe HCN effects on it, the much shorter processes should be taken into account. So, we developed also a model incorporating two gating variables, fast activation and slow inactivation (similar to m and h in HH equations).
\\

{\it  Stem cell model with activation and inactivation (Stem-mh)} is:
 
\begin{subequations}
\label{Stem-mh}
\begin{eqnarray}
d E_s /dt = -N  \sigma_f  \,g_1^s g_2^s (E_s -E_r ) /C_s \\
d g_i^s /dt  =  (  \bar{g}_i^s - g_i^s)  /  \tau_i^s(E_s),\,\,\ i=1,2            \end{eqnarray}
\end{subequations}

$E_s$ is the stem cell membrane potential, mV.
$g_1^s, g_2^s$  are activation and inactivation gating variables  
(notation $m, h$ are used in Hodgkin-Huxley eqns\cite{HH}). 
$\tau_1^s, \tau_2^s$ are activation and inactivation time constants, 
Fig.\ref{g &tau_mh}.  \\
 
\vspace{.2cm}

\vspace{.2cm}
{\bf Different types of myocytes }

We need to investigate interaction of stem cells with myocytes of different types. No published models are sufficient for this (every model describes a specific type of myocyte). For this reason, we will content ourselves with an existing model, which we will modify to reproduce qualitatively the various types of myocytes.
  
In this article, we work with several types of myocyte:  
ventricular adult myocyte (with high threshold $I_{osc}$ for inducing oscillations); 
self-oscillating neonatal type myocyte and several transitional types. 
We obtain models for these myocytes from a model for ventricular myocytes, using different values of expression levels of $I_{K1}$ current and re-scaling the time constants to keep the action potential duration (APD) identical for all myocytes.

So, $I_{K1}$ is expressed as:

\begin{eqnarray}
I_{K1} = k'   \, \,   I^1_{K1} 		
\label{kcoef}
\end{eqnarray}

where $k'$ is the number of $I_{K1}$ channels per myocyte and $I^1_{K1}$ is the current through one $I_{K1}$ channel (averaged in time). 

Our models were derived from the  
Beeler-Reuter model (BR) \cite{BR}, which describes the membrane  potential, 
$E_m$, gating variables $g_i^m$ controlling ionic chanels and Calcium concentration $[Ca]$:

\begin{subequations}
\label{BR}
\begin{eqnarray}
\partial_t E_m &=& - I_{ion}(E_m, {g_i^m})/C_m + D \nabla^2 E_m
\\
\partial_t {g_i^m} &=& ( { \bar{g}_i^m} - g_i^m)  /  \tau_i^m(E_m),\,\,\ i=1,6
%\label{BRg}
\\
\partial_t{[Ca]}&=&-10^{-7} i_{Ca}+ 0.07(10^{-7} - [Ca])
%\label{Ca}
\end{eqnarray}
\end{subequations}

Here $i_{Ca}=g_{Ca}g_5^m g_6^m(E_m-E_{Ca})$ , $g_{Ca}=0.09mS/cm^2 $ and $E_{Ca}=-82.3-13.0287\ln [Ca]$.
The diffusion term $D$ in Eq.\ref{BR} represents the electric coupling between cells: $D\sim10^{-3} cm^2/mSec$. 
The model was calculated in one and two dimensional cases with $dx=0.1 mm and dt=0.02 mSec$.

A myocyte with $k'$= 1400 $I_{K1}$ channels per myocyte corresponds to the original BR model. We will use also dimensionless values $k = k' / 1400$.
 The excitation threshold is $I_{exc}= 2  pA/pF$ and the oscillation threshold $I_{osc} = 2.01  pA/pF$. The myocytes are auto-oscillatory for $k'<\sim 320$ ($k<\sim 0.23$) and are excitable for $k'>\sim 320$.  

 When $k'$ approaches from below to the critical value $k'_c\sim 320$, the amplitude of the oscillation remains the same, while the period $T_{osc}$ of the oscillation diverges logarithmically $T_{osc}\sim ln(k_c - k), k<k_c$. 
  This indicates the change from oscillatory to excitable but quiescent behavior occurs through a homoclinic bifurcation \cite{bif},\cite{bifurcKurata}).

 \begin{figure}[h]
\includegraphics[width=3.2in]{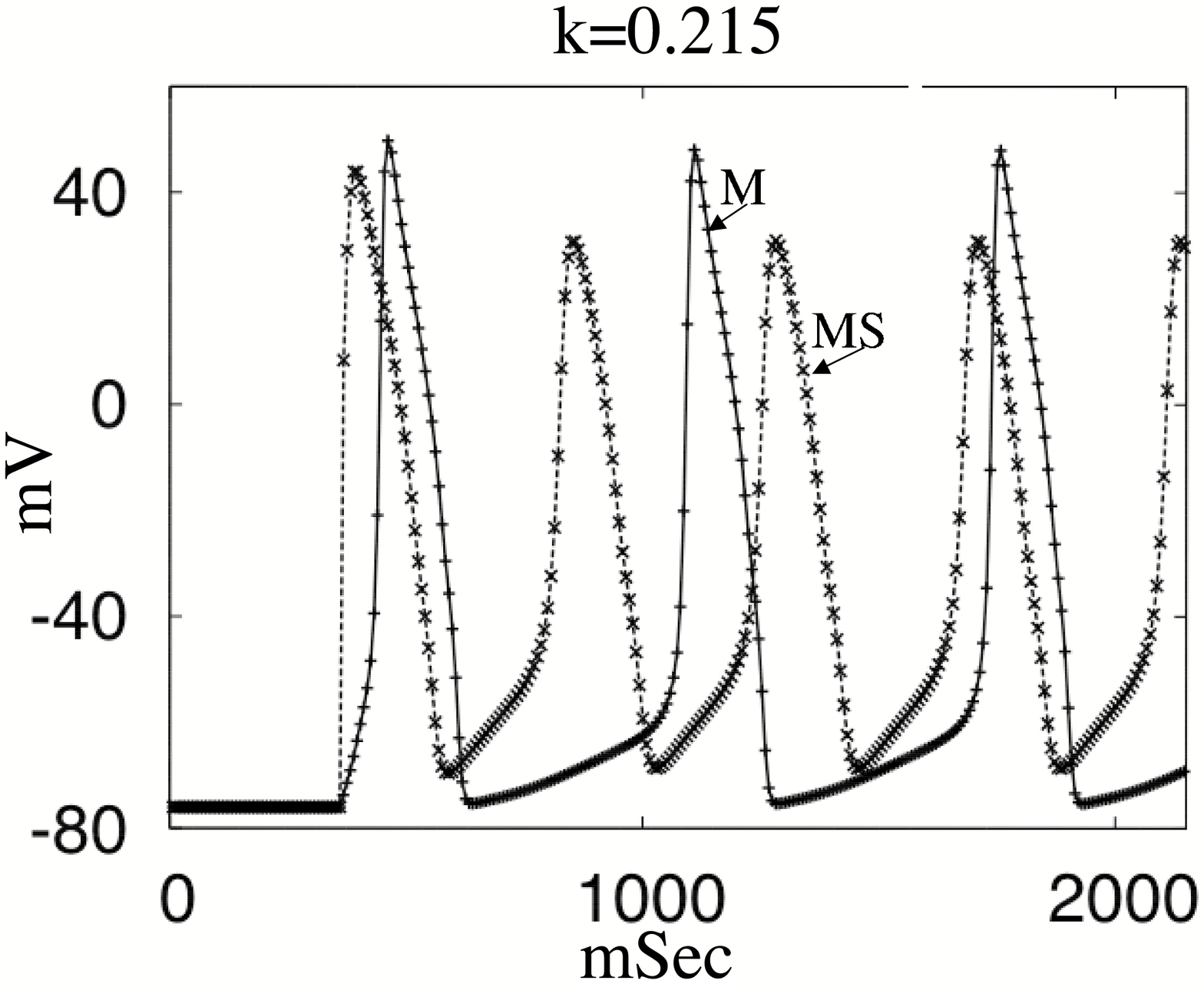}
\caption[]{ {\bf Decrease of oscillations period in neonatal myocytes  by HCN2 transfected stem cells.}
A cell culture consisting of neonatal self-oscillating cells was simulated.
A record far from the pacemaker region: M- myocytes alone; MS- stem cells are added.
A one dimensional simulation: 201 myocytes with $k'$= 300 $I_{K1}$ channels per myocyte, myocytes alone oscillate  with period  $T_0$=650ms. Stem cells with $N=10^5$ pacemaker channels per stem cell, n= 1000 gap junctions are added to the pacemaker region $l=4.3 mm$ long. They decreased the period to 425 ms.  
}
\label{neonatal}
\end{figure}

\begin{figure}[h]
\includegraphics[width=3.9in]{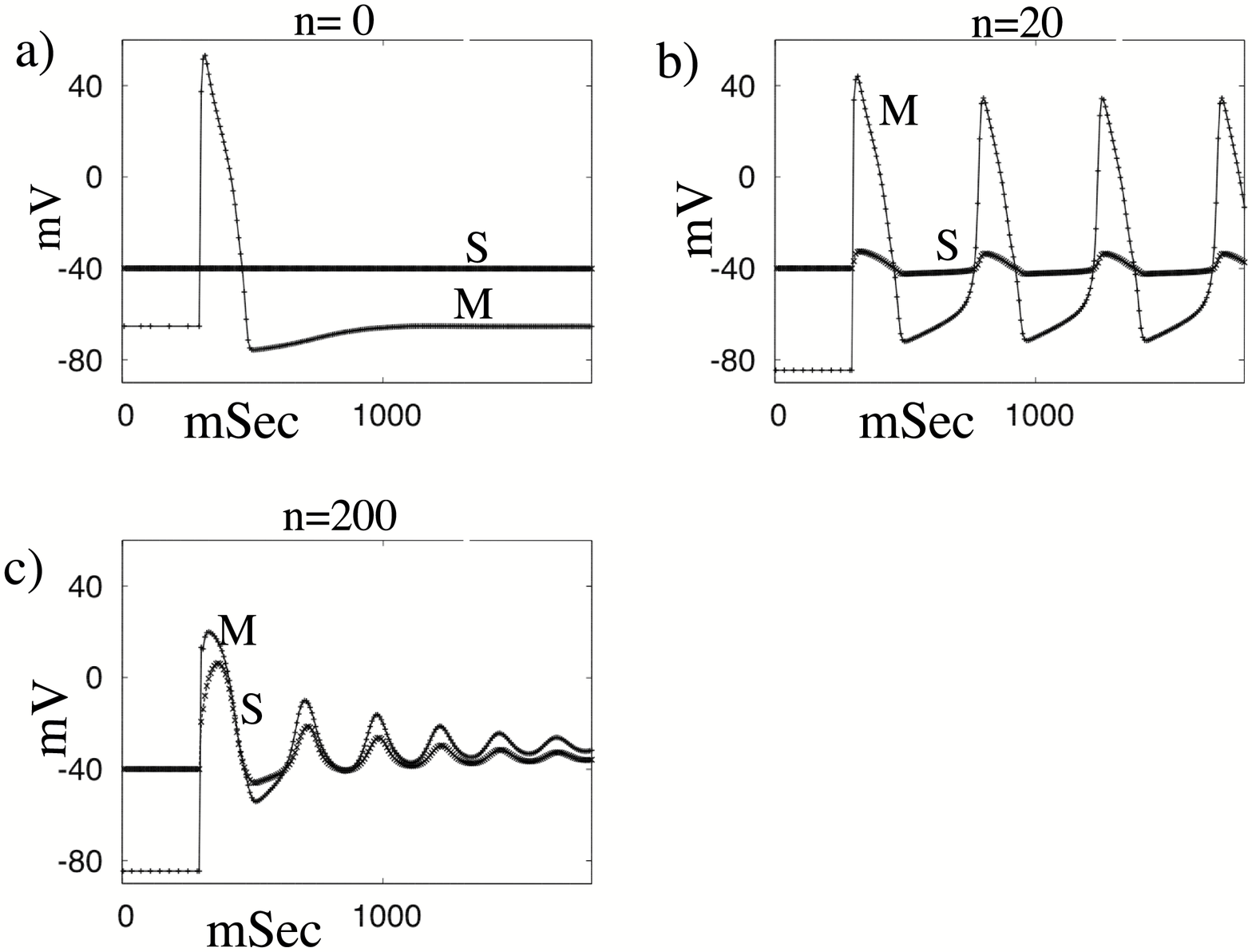}
\caption[]{  {\bf Induction of oscillations by a stem cell connected to a myocyte. }
Myocyte and stem cell potentials are noted as M and S. Above each image the number of gap junctions $n$ per cell is shown.
a) cells are not connected (n=0), the  myocyte generates one AP only. 
b) oscillations in the cell pair.
c) disappearance of oscillations when the cell connection is too strong.
 Parameters: the number of pacemaker channels per stem cell $N= 1.2\times10^5$ (see eq.\ref{stemconnect}a), the number of $I_{K1}$ channels per myocyte $k'=326$ (see eq.\ref{kcoef}). 
} 
\label{nN}
\end{figure}

\vspace{.4cm}

{\bf Cell pair model} 

  A "cell pair", consisting of a myocyte and a stem cell coupled
together is described by the following set of equations :

\begin{figure}
\includegraphics[width=3.2in]{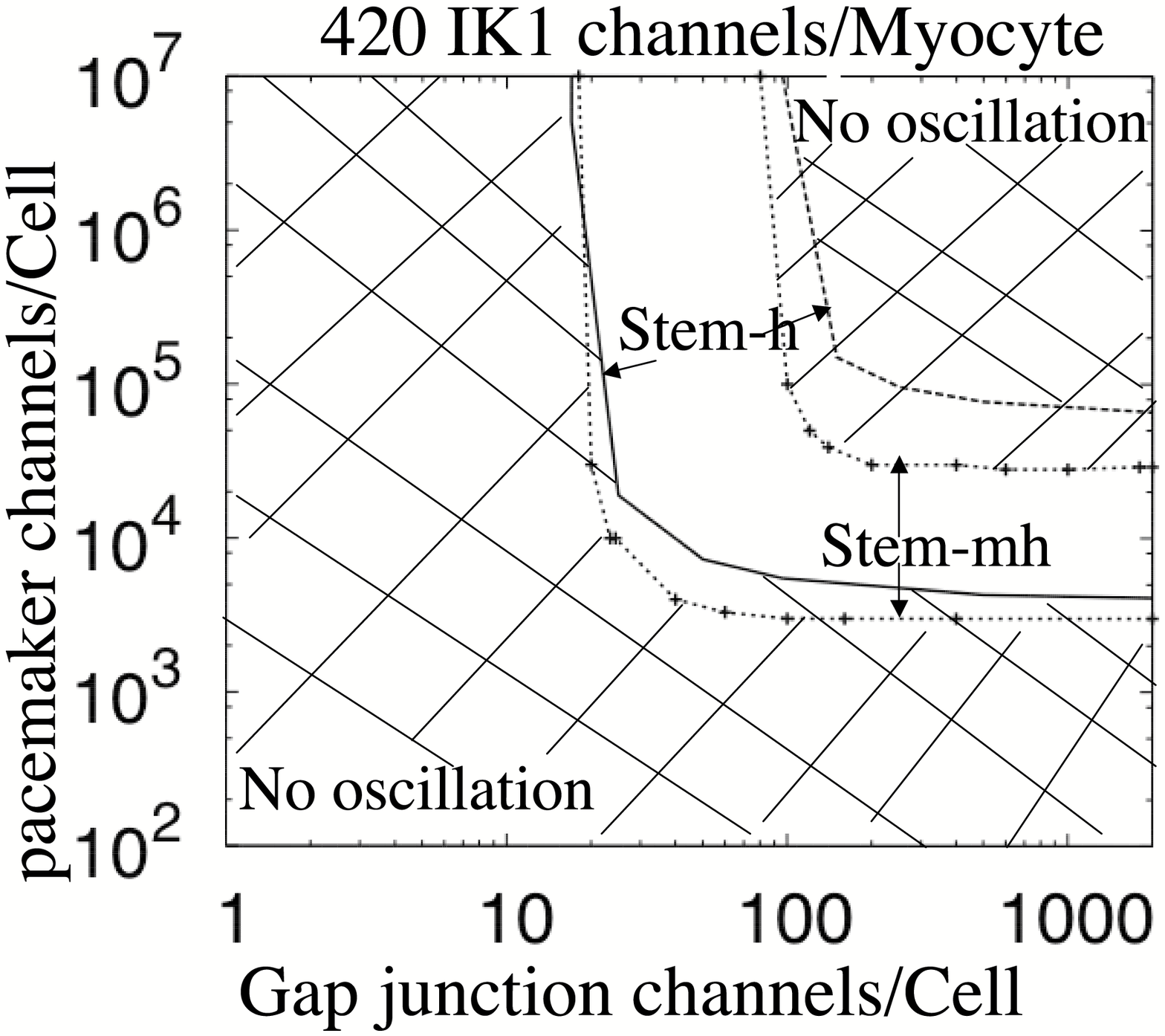}
\caption[]{ {\bf Oscillations phase plane for a cell pair}.
 Boundaries of oscillation regions for two models are shown: a model  with inactivation only (model Stem-h, Eq.\ref{Stem-h}) and a model with activation and inactivation (model Stem-mh, Eq.\ref{Stem-mh}).    Note that in both models, high expression levels of gap junctions or pacemaker channels are not needed to induce oscillations and may even destroy them.
}  
\label{zone}
\end{figure}

\begin{subequations}
\label{stemconnect}
\begin{eqnarray}
\partial_t E_s &=& -N  \sigma_f  \, g_2^s (E_s + 40) /C_s+\nonumber
 \\&&n \sigma_g  (E_m - E_s)/C_s
\\
\partial_t E_m &=& - I_{ion}(E_m, {g_i^m})/C_m +\nonumber 
\\&&n \sigma_g (E_s - E_m)/C_m 
\label{myoconnect}
\\
\partial_t {g_i^m} &=& F( {g_i^m},E_m )/\tau_i^m (E_m),\,\,\ i=1,...,6
\label{BRg}
\\
\partial_t{[Ca]}&=&-10^{-7} i_{Ca}+ 0.07(10^{-7} - [Ca])
\label{CaCouple}
\\
d g_2^s /dt &=& F( g_2^s,E_s )/\tau_2^s(E_s)
\label{stemgc}
\end{eqnarray}
\end{subequations}
 
where $n$ is the number of gap junction channels per cell, 
$\sigma_g$ is the conductivity of a single gap junction channel (50 $pS$). 
Term $ n \sigma_g  (E_m - E_s)$ in Eq.\ref{stemconnect}a describes current flowing from a myocyte to a stem cell.

Here, we described the stem cell by the simplified model Eq.\ref{Stem-h}. The implementation of the full model Eq.\ref{Stem-mh} leads to a 
very similar system.

\vspace{.4cm}
{\bf Cardiac tissue with incorporated stem cells }

We suppose that stem cells are injected to a pacemaker region only and are connected to myocytes. The tissue is described 
by Eqs.\ref{BR}, \ref{stemconnect}.
Inside the pacemaker region, to Eq.\ref{BR}a, the term 
$n \sigma_g  (E_s - E_m)$ is added, similar to Eq. \ref{stemconnect}b. 
Numerically, we considered both a one dimensional (1D) fiber and a two dimensional (2D) tissue.

\vspace{.4cm}

{\Large {\bf Increase of oscillation frequency in  self-oscillatory myocytes.} }

In experiments \cite{IraC1}, stem cells added to cultured neonatal myocytes that are self-oscillatory. The stem cells decreased the oscillation period of these myocytes from $T \sim 660$ to $T \sim 370 $ms. This phenomenon is easily reproduced by the model, both in a cell pair and in a cardiac tissue, Fig.\ref{neonatal}.

This decreasing of oscillation frequency by HCN genes is a robust phenomenon. In addition, it can be observed at arbitrary values of parameters; only the amount of the period decrease is parameter dependent. 

 A large interval of number of pacemaker channels per cell $N$ and number of gap junction channels per cell $n$ can be easily investigated. In todays experiments, only a part of this interval is  accessible. 

\vspace{.4cm}
{\Large {\bf Induction of oscillations in quiescent myocytes.}}

{\bf Induction of oscillations in a cell pair}

Induction of oscillations by a stem cell connected to a myocyte is shown in Fig.\ref{nN}.
When the cells are not connected (Fig.\ref{nN}a, n=0), the myocyte 
can generate one AP only. 
When the cells are connected (Fig.\ref{nN}b, n=20), a pacemaker activity can be initiated in the cell pair. A diastolic depolarization is observed in the myocytes, and it generates large amplitude periodic APs. 

This result confirms the conjecture, based on results of experiments with tissues  \cite{IraPrivate}, \cite{IraC1}, that a stem cell {\it hMSC} transfected with HCN2 gene can induce oscillations 
in a cardiac cell.  

\vspace{.2cm}

{\it Disappearance of oscillations }

When the cells are too strongly connected (Fig.\ref{nN}c, n=200), the 
pacemaker activity decays: the diastolic potential shifts closer to the resting potential of the stem cell ($E_r \sim -40 mV$) and the sodium channels are inactivated. This is not a sharp termination of oscillations: the oscillation amplitude decreases smoothly as the number of gap junctions $n$ increases. When the amplitude of generated periodical AP's is too small, they will be unable to propagate in cardiac tissue.

In Fig.\ref{zone} the position of "disappearance of oscillation" 
boundary was chosen at parameter values where the upstroke of the AP decreases below $\sim 0$ mV and the amplitude of oscillations becomes inferior $\sim 50$ mV. It would be more accurate to determine when the propagation will disappear using cardiac tissue models.

\vspace{.25cm}

{\it Robustness of oscillations.}  

As we observed from the example above, oscillations in excitable cardiac cells can be induced not for all parameter values.

In Fig.\ref{zone} phase plane for a cell oscillation pair is shown.
It is seen that oscillations are not observed at low values of number $N$ of pacemaker channels per cell and number $n$ of gap junction channels per cell.
When n, N are increasing, oscillations arrive as a jump, with full amplitude but with a very high oscillations period (several seconds). The period decreases and becomes normal when N,n continue increasing. It is the same homoclinic bifurcation \cite{bif} that describes the transition from oscillatory to excitable but quiescent behavior when expression level of $I_{K1}$ is increased.

At the contrary, when oscillations disappear in a cell pair, there is no jump: amplitude of oscillations slowly decrease while values of n, N increasing too large. 

Fig.\ref{zone} shows the oscillation boundaries for two models: the  model comprising inactivation gating variable only as in majority of the descriptions published (model Stem-h), and the model comprising  both gating variables, activation and inactivation (model Stem-mh). 
Since the ratio $\epsilon= \tau_1^s/ \tau_2^s \ \ll 1$ (Fig. \ref{g &tau_mh}b), it is natural to expect that a simplified model with inactivation only will give results close to the full model. As illustrated in Fig.\ref{zone} the results are qualitatively but not quantitatively similar.

As seen in Fig.\ref{zone} the induced oscillations are robust: they exist for a wide range of parameters: number of gap junctions per cell $n$ and number of pacemaker channels per cell $N$. This is important to consider for constructing a biological pacemaker.

\vspace{0.2cm}
\textsl{Cell pair experiments}. Induction of oscillations in a cell pair experiment was not demonstrated yet. What is the reason? A possible explanation : Expression of gap junctions in a cell pair ($\sim$ 100 gap junctions/cell) is 1-2 orders of magnitude less than in a normal cardiac tissue. Then, a path to induce oscillations in a cell pair might be increasing the expression of gap junctions. 

As Fig.\ref{zone} show, approximately 100 gap junctions per cell is well enough for induction of oscillations. The models show that :

- an experimental path intended to increase the expression of gap junctions will not give positive results.

- selecting myocytes with not high level of expression of $I_{K1}$ will result in oscillations.

{\bf Conclusions} 

1. Mathematical models of stem cells (hMSC) transfected with HCN2 genes and connected to myocytes were developed. This opens a way to investigate the induction of oscillation and its robustness in a cell pair, in cell culture and in the three dimensional cardiac tissue. 

2. The mathematical results confirm the conjecture, based on  experiments with tissues, that stem cells can induce oscillation even in myocytes that do not self- oscillate. The oscillations are robust and well suited for creation of biological pacemakers (Fig.\ref{zone}).

3. To induce oscillations in a cell pair, there is no need to increase expression levels of gap junctions or of HCN genes (Fig.\ref{zone}). 

4. The models have a potential to eliminate a part of time consuming genetical and electrophysiological experiments needed for developing a biological pacemaker. They also permit to estimate the effects of expression levels that are not achieved in the experiments yet (Figs. \ref{zone}). 

{\bf Acknowledgment} 
The authors acknowledge the helpful discussions with Claudia Altomare, M. Boyet, I. S. Cohen, Halina Dobrzynski, F. Fenton, M. Mangoni, H. Ohmori, D. Pazo, S. Takagi, C. Ulens.
%references


\begin{references}
 
\bibitem{porcine2001} Edelberg JM, Huang DT, Josephson ME, Rosenberg RD. Molecular enhancement of porcine cardiac chronotropy. Heart 2001;86:559Ð 562.

\bibitem{Ira2001} Qu J, Barbuti A, Protas L, Santoro B, Cohen IS, Robinson RB. HCN2 overexpression in newborn and adult ventricular myocytes: distinct effects on gating and excitability. Circ Res. 2001 Jul 6;89(1):E8-14.

\bibitem{Marban2002Nature}  Miake J, Marb‡n E, Nuss HB. Gene therapy: biological pacemaker created by gene transfer. Nature. 2002;419:132Ð133.

\bibitem{Rosen1} Plotnikov AN, Sosunov EA, Qu J, Shlapakova IN, Anyukhovsky EP, Liu L, Janse MJ, Brink PR, Cohen IS, Robinson RB, Danilo P, Rosen MR. Biological Pacemaker Implanted in Canine Left Bundle Branch Provides Ventricular Escape Rhythms That Have Physiologically Acceptable Rates.  Circulation. 2004;109:506-512.

\bibitem{Rosen}  Rosen M, 15th annual Gordon K. Moe Lecture. Biological pacemaking: in our lifetime?
Heart Rhythm. 2005 Apr;2(4):418-28.

\bibitem{IraC1} Potapova I, Plotnikov A, Lu Z, Danilo P, Jr, Valiunas V, Qu J, Doronin S, Zuckerman J, Shlapakova IN, Gao J, Pan Z, Herron AJ, Robinson RB, Brink PR, Rosen MR, Cohen IS. Human Mesenchymal Stem Cells as a Gene Delivery System to Create Cardiac Pacemakers. Circulation.  2004;94:952-959.

\bibitem{IraPrivate} Cohen IS, private commnication, 2. 20. 2005

\bibitem{IraC2} Valiunas V, Doronin S, Valiuniene L, Potapova I, Zuckerman J, Walcott B, Robinson RB, Rosen MR, Brink PR, Cohen IS. Human mesenchymal stem cells make cardiac connexins and form functional gap junctions. 2004.J Physiol 555.3 pp 617-626.

\bibitem{IraC3} Rosen MR, Brink PR, Cohen IS, Robinson RB. Genes, stem cells and biological pacemakers. Cardiovascular research. 2004; 64 : 12-23.

\bibitem{kehat} Kehat I, Khimovich L, Caspi O, Gepstein A, Shofti R, Arbel G, Huber I, Satin J, Itskovitz-Eldor J, Gepstein L. Electromechanical integration of cardiomyocytes derived from human embryonic stem cells. 2004; doi:10.1038/nbt1014.

\bibitem{Qu} Qu J, Kryukova Y, Potapova IA, Doronin SV, Larsen M, Krishnamurthy G, Cohen IS, Robinson RB. MiRP1 modulates HCN2 channel expression and gating in cardiac myocytes.
J Biol Chem. 2004 Oct 15;279(42):43497-502.

\bibitem{BR} Beeler GW, Reuter H. Reconstruction of the action potential of ventricular myocardium fibers. J Physiol.1977; 268:177-210. 

%2

\bibitem{modelKurata} Kurata Y, Hisatome I, Imanishi S, Shibamoto T.
Dynamical description of sinoatrial node pacemaking: improved mathematical model for primary pacemaker cell.
Am J Physiol Heart Circ Physiol. 2002 Nov;283(5):H2074-101. 
 
 
\bibitem{bifurcKurata} Kurata Y, Hisatome I,  Matsuda H, Shibamoto T.
Dynamical Mechanisms of Pacemaker Generation in IK1-Downregulated Human Ventricular Myocytes: Insights from Bifurcation Analyses of a Mathematical Model 
Biophys. J. 2005 89: 2865-2887.

\bibitem{ZhangBoyett}Zhang H, Holden AV, Kodama I, Honjo H, Lei M, Varghese T, Boyett MR. Mathematical models of action potentials in the periphery and center of the rabbit sinoatrial node.
Am J Physiol Heart Circ Physiol. 2000 Jul;279(1):H397-421.

\bibitem{Marban05}Azene EM, Xue T, Marban E, Tomaselli GF, Li RA. 	Non-equilibrium behavior of HCN channels: insights into the role of HCN channels in native and engineered pacemakers.
Cardiovasc Res. 2005 Aug 1;67(2):263-73.


\bibitem{Gepstein} Gepstein L. Derivation and potential applications of human embryonic stem cells. Circ Res. 2002;  91:866Ð876. 

\bibitem{HH} Hodgkin AL, Huxley F. A quantitative description of memebrane current and it's application to conduction and excitation in nerve. J Physiol. 1952;  117:500-544.

\bibitem{Valiu} Valiunas V, Weingart R, Brink PR. Formation of heterotypic gap junction channels by connexins 40 and 43. Circ Res. 2000;86:e42Ðe49.

\bibitem{Moosmang} Moosmang S, Stieber J, Zong X, Biel M, Hofmann F, Ludwig A. Cellular expression and functional characterization of four hyperpolarization-activated pacemaker channels in cardiac and neuronal tissues. Eur J Biochem. 2001;268:1646Ð1652. 

\bibitem{DiFrancesco} Accili EA, Proenza C, Baruscotti M, DiFrancesco D. From Funny Current to HCN Channels: 20 Years of Excitation. J Physiol. News Physiol Sci 17: 32-37, 2002. 

\bibitem{Yu} Yu H, Wu J, Potapova I, Wymore RT, Holmes B, Zuckerman J,  Pan Z, Wang H, Shi W, Robinson RB, El-Maghrabi MR, Benjamin W, Dixon J, McKinnon D, Cohen IS, Wymore R. MinK-Related Peptide 1, A $\beta$ Subunit for the HCN Ion Channel Subunit Family Enhances Expression and Speeds Activation. Circulation Research. 2001;88:e84.

\bibitem{Moore}Moore LK, Burt JM. Gap junction function in vascular smooth muscle: influence of serotonin.  Am J Physiol. 1995 Oct;269(4 Pt 2):H1481-9.

%\bibitem{}
%\bibitem{}
%\bibitem{}
%\bibitem{}

\bibitem{bif} Pumir A, Arutunyan A, Krinsky  V, Sarvazyan N.
Genesis of ectopic waves: role of coupling, automaticity and heterogeneity.
Biophys. J, to appear.
\bibitem{Bender} Bender  CM, Orszag SA. Advanced mathematical
methods for scientists and engineers, Mac Graw Hill, Singapore (1984).

\bibitem{Abramowitz} Abramowitz M, Stegun IA. Handbook of mathematical functions. Dover, 1972.

\end{references}
\end{document}